\documentclass{article}

\usepackage{PRIMEarxiv}

\usepackage[utf8]{inputenc} 
\usepackage[T1]{fontenc}    
\usepackage{hyperref}       
\usepackage{url}            
\usepackage{booktabs}       
\usepackage{amsfonts}       
\usepackage{nicefrac}       
\usepackage{microtype}      
\usepackage{lipsum}
\usepackage{fancyhdr}       
\usepackage{graphicx}       
\graphicspath{{media/}}     

\title{Random quasi-phase-matching for pulse characterization from the near to the long wavelength infrared}

\author{
  Brandin Davis, Tobias Saule, Carlos A. Trallero-Herrero \\
  Department of Physics\\
  University of Connecticut \\
  Storrs, Connecticut 06269, USA\\
  \texttt{carlos.trallero@uconn.edu} \\
}

\begin{document}
\maketitle

\begin{abstract}
Experiments requiring ultrafast laser pulses require a full characterization of the electric field to glean meaning from the experimental data. Such characterization typically requires a separate parametric optical process. As the central wavelength range of new sources continues to increase so too does the need for nonlinear crystals suited for characterizing these wavelengths. Here we report on the use of poly-crystalline zinc selenide as a universal nonlinear crystal in the frequency resolved optical gating characterization technique from the near to long-wavelength infrared. Due to its property of random quasi-phase-matching it's capable of phase matching second-harmonic and sum-frequency generation of ultra-broadband pulses in the near and long wavelength infrared, while being crystal orientation independent. With the majority of ultra-fast laser sources being in this span of wavelengths, this work demonstrates a greatly simplified approach towards ultra-fast pulse characterization spanning from the near to the long-wavelength infrared. To our knowledge there is no single optical technique capable of such flexible capabilities.
\end{abstract}


\section{Introduction}
Since the first demonstration of sub picosecond pulsed lasers, non-electronic based techniques were required to fully characterize these ultra-fast sources of laser light. Many techniques in this regard have been developed, including optical autocorrelation, interferometric autocorrelation, spectral phase interferometry for direct electric-field reconstruction (SPIDER) \cite{Iaconis1998}, grating-eliminated no-nonsense observation of ultrafast incident laser light e-fields (GRENOUILLE) \cite{OShea2001}, dispersion scan (d-scan) \cite{Miranda2012}, stereo above threshold ionization \cite{Sayler2011}, and electro-optic sampling (EOS) \cite{Wu1996} to name a few. One of the first and still most popular method for characterizing ultra-fast pulses of light is second-harmonic generation frequency resolved optical gating (SHG-FROG)\cite{Trebino1997}. In this method the pulse to be measured is used to characterize itself through self gating in a nonlinear medium, producing a second-harmonic spectrogram (higher order harmonics, such as the third, can also be used but this paper will focus on SHG) as a function of the delay of the gate, which can be used to algorithmically retrieve amplitude and phase information about the pulse itself. A generalization of this technique, known as cross-correlated frequency resolved optical gating (X-FROG)\cite{Dudley2002}, involves overlapping two different input pulses (the gating pulse which is measured through SHG-FROG and the unknown pulse) and measuring their sum-frequency spectrogram, allowing for reconstruction of the originally unknown pulse. Often X-FROG is used to characterize a pulse whose second-harmonic isn't measurable with commonly used silicon based spectrometers. For example in \cite{Wilson2019} an X-FROG measurement of a femtosecond long-wavelength infrared (LWIR, 7-14~$\mu$m) pulse is performed using a gate pulse ten times shorter in center wavelength.

In both FROG and X-FROG, the nonlinear crystal being used needs to be transmissive for all wavelengths involved and have phase-matching properties to support the bandwidth of the nonlinear process. These requirements can be severely limiting, especially in the case of X-FROG where one may be attempting to characterize a pulse in the LWIR with a gating pulse in the near-infrared (NIR, 780-2500~nm). While a typical SHG-FROG or X-FROG setup is optically simple, great care must be taken when choosing the nonlinear crystal, for the reasons stated above, to insure an accurate reconstruction of the pulse. In the visible to NIR spectral range there are several crystals that can be used for FROG/XFROG, with $\beta$-BaB2O4 (BBO) one of the most commonly used. On the other hand, crystals in the mid to long wavelength infrared are often costly, difficult to grow, and sometimes unavailable altogether. Additionally, these nonlinear crystals are generally limited to narrow spectral regions such as the MWIR, NIR, or LWIR, making them very application specific. For example, gallium selenide, which is capable of phase matching from 2$\mu$m-13$\mu$m, can only do so by tuning the phase matching angle of the crystal over nearly 32 degrees. Such large changes in the phase matching angle can bring its own host of issues, some of which are highlighted in \cite{Tsubouchi:09}. For these reasons there have been few reports on the optical characterization of LWIR femtosecond pulses using the X-FROG technique with nonlinear crystals \cite{Wilson2019,Heiner:19,Heiner:20,Tsubouchi:09,Liang2017} with some choosing to do X-FROG in gas media using the four wave mixing technique due to gas having much less dispersion and thus better phase matching properties, though at the cost of increased complexity \cite{Lanin:14, 10.1117/12.2507460, Nie2020}. 

Here we report on results which show poly-crystalline zinc selenide (pc-ZnSe) meets all requirements for a universal crystal for pulse characterization from the NIR to the LWIR due to its large transmission window and random quasi-phase-matching properties (RQPM), which negate the need for tuning of the phase matching angle of the crystal \cite{Baudrier-Raybaut2004,Kawamori2019,10.1117/1.3257266,doi:10.1063/1.5054390,Zhang:19,Schunemann:20}. We compare SHG-FROG and X-FROG measurements made with pc-ZnSe, AgGaS$_2$ (AGS), and BBO to demonstrate its ability to characterize ultra-broadband pulses with center wavelengths from 1~$\mu$m to 10~$\mu$m.

\section{Experiment}
\begin{figure}[h!]
\centering\includegraphics[width=0.75\textwidth]{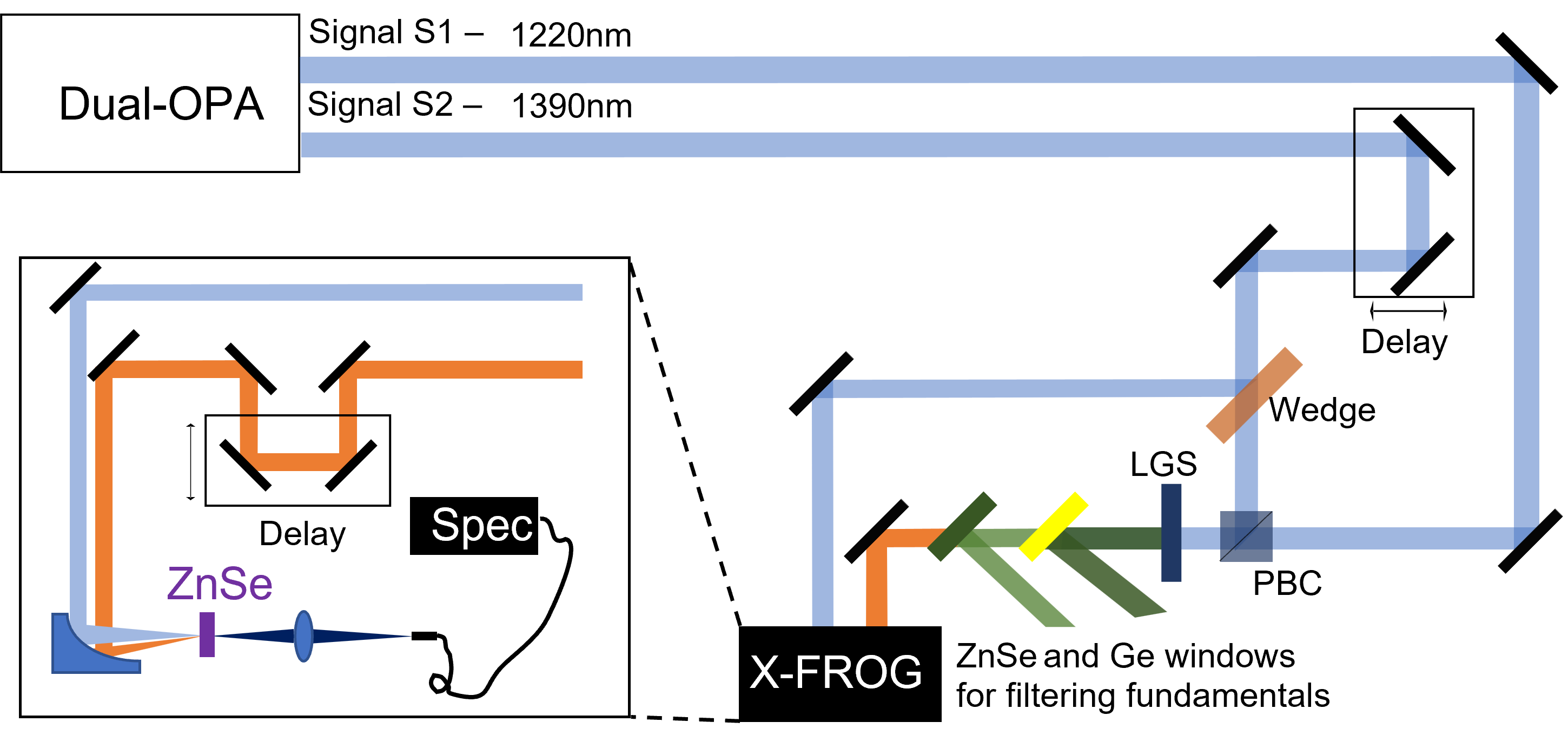}
\caption{Experimental setup for LWIR generation and characterization with X-FROG. The two output signals S$_1$ and S$_2$ of a dual-OPA are overlapped in space and time in the crystal LGS to produce wavelength tunable LWIR. The signals are then filtered with two anti-reflection coated windows and a small amount of the weak signal is picked off with a uncoated calcium flouride wedge to be used as the gate for X-FROG of the LWIR. The inset shows a sketch of the X-FROG setup. Both beams are focused with an f=6~inch off-axis parabolic mirror. The FROG setup is identical but with both arms of the interferometer being degenerate.}
\label{fig:Figure 1}
\end{figure}

Our experimental setup starts with a Ti:Sapphire laser from Continuum USA, capable of delivering 18~mJ of energy per pulse, 35~fs in duration, at 1~kHz repetition rate, and 800~nm center wavelength. The laser pumps a high energy, asymmetric dual-OPA\cite{Davis2021} that generates two signals of tunable wavelength and a combined energy of 3~mJ. These signals are overlapped in the crystal lithium gallium sulfide (LGS, 1.2~mm thick, $\theta$ = 51.8) and produce 25~$\mu$J of tunable LWIR \cite{Petrov2004, doi:10.1126/sciadv.aaq1526, Qu:19}  (see Fig. \ref{fig:Figure 1}). Here we present results with the signals S$_1$ and S$_2$ at wavelengths, $\lambda_1 = 1220$nm and $\lambda_2=1390$~nm respectively. After generation, S$_1$ and S$_2$ are filtered out first with a zinc selenide window, then with a germanium window (Edmund Optics, stock number 68-512 and 83-349 respectively) both coated to transmit the LWIR and reflect the signals in such a manner as to avoid two-photon absorption which can cause unwanted absorption of the generated LWIR \cite{Wilson2019} in Ge. The X-FROG gating field is obtained by picking off a small percentage of S$_2$ which is characterized through SHG-FROG using BBO and pc-ZnSe (In both cases an Ocean Optics HR4000 spectrometer was used). The BBO crystal is 500~$\mu$m thick, cut for phase matching of the second-harmonic of 1600~nm. Our pc-ZnSe  is a 1mm thick uncoated window from Edmund Optics (stock number 39-394). All future references to pc-ZnSe in this paper refer to this window. It should be noted that while thinner samples would impart less GVD on the input pulses they also offer less overall conversion efficiency which increases linearly with length\cite{Baudrier-Raybaut2004}. For instance at 9$\mu$m the group delay dispersion imparted by our 400$\mu$m thick AGS crystal is -560fs$^2$, while the pc-ZnSe crystal imparts -680fs$^2$. Thus we found the 1mm thick pc-ZnSe to be the best compromise for this study. The pc-ZnSe FROG reconstructed S$_2$ field is then subsequently used in the X-FROG reconstructions of the LWIR pulses, where AGS and pc-ZnSe are compared. Both the SHG-FROG and X-FROG setup are all reflective, using Ag mirrors that support the entire bandwidth of all involved pulses and off axis parabolic mirrors for focusing (X-FROG spectrograms were taken with a HORIBA CP140 spectrograph and Sensors Unlimited SU-LDV-512LDB NIR camera combination). FROG traces of the idler (1885nm) were also compared using BBO and pc-ZnSe which showed good agreement but will be omitted for brevity. For all pulse reconstructions we use the method of projections \cite{DeLong1994} as implemented in the open source repository \cite{femtosoft}. In all cases use of the noise subtraction features (full spectrum, edge, and cleanup pixels) in the cited software were used.

\begin{figure}[h!]
\centering
\includegraphics[width=1.0\textwidth]{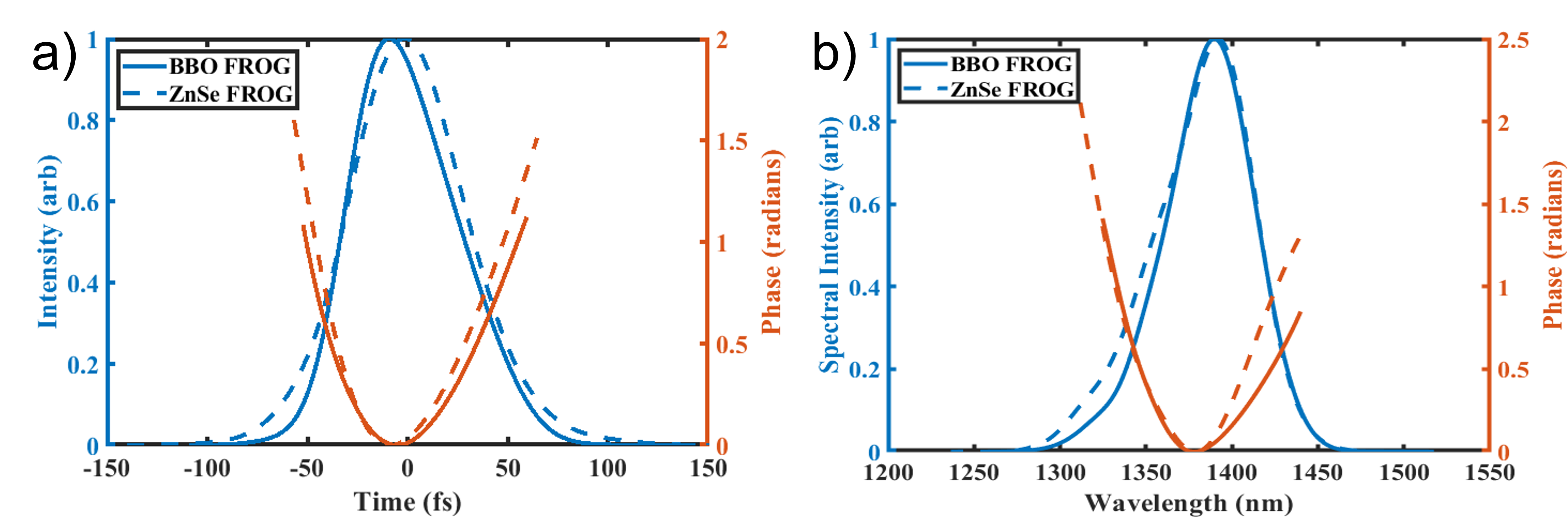}
\caption{SHG-FROG reconstruction of the OPA signal S$_2$ (the gate pulse for the X-FROG) using both BBO and pc-ZnSe. In a) reconstructed amplitudes and phases in the time domain compared, b) reconstructed amplitudes and phases in the spectral domain compared. In all plots blue is the normalized amplitude and orange is the phase. The rms phase deviations are 0.14~rad and 0.17~rad respectively. Both FROG reconstructions used a grid size of 2048~x~2048. The FROG error was 0.0018 for BBO and 0.0036 for pc-ZnSe.}
\label{fig:Figure 2}
\end{figure}

\section{Results}

Figure \ref{fig:Figure 2} shows field retrievals of the S$_2$ field with BBO or pc-ZnSe crystals in the SHG-FROG setup. As seen in the figure, field reconstruction when using pc-ZnSe very closely matches that of the reconstruction when using BBO. In the case of BBO the full-width at half maximum (FWHM) time duration was reconstructed to be 63~fs, while that produced with pc-ZnSe was 67~fs, while the measured pulse durations (taken from the autocorrelation data of the FROG spectrograms) were 62fs in the case of BBO and 69.5fs for pc-ZnSe, showing good agreement between measured and retrieved pulse durations. The retrieved spectral FWHM was 59nm for BBO and 65nm for pc-ZnSe. The slightly longer duration with pc-ZnSe is due to ZnSe having 330~fs$^2$ more group delay dispersion than BBO at the signal wavelength. The root mean square (RMS) deviation between the temporal phase and spectral phase of both measurements is 0.14~rad and 0.17~rad respectively.  Additionally, in the case of pc-ZnSe, the produced second-harmonic spectrogram was independent of the polarization orientation with respect to the crystal.

\begin{figure}[h!]
\centering\includegraphics[width=0.9\textwidth]{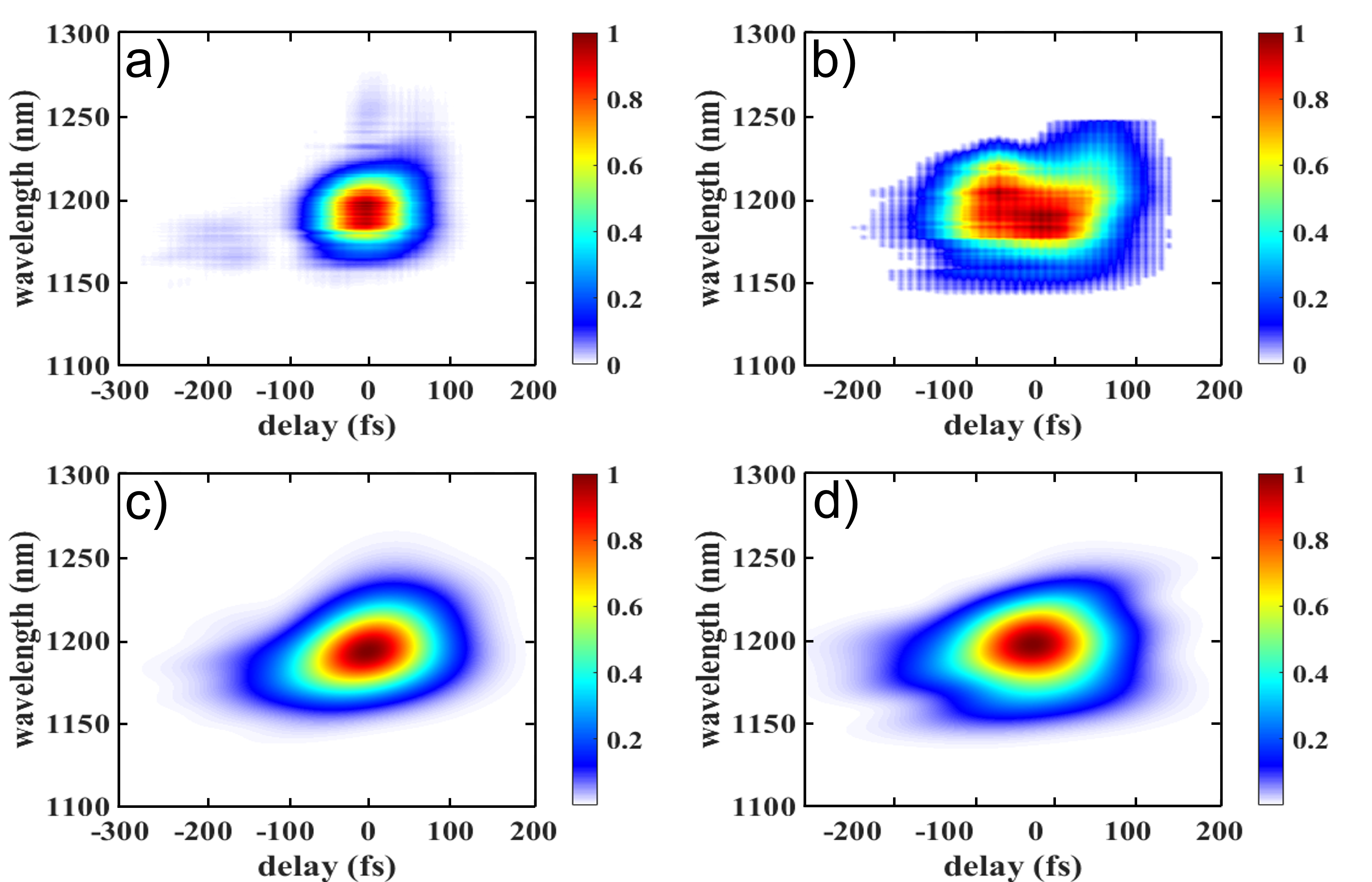}
\caption{Raw and retrieved X-FROG traces of the LWIR. (a) Raw spectrogram and (c) reconstructed trace using crystal AGS. (b) Raw spectrogram and (d) reconstructed trace using pc-ZnSe. In all measurements S$_2$, as shown in Fig. \ref{fig:Figure 2} with pc-ZnSe was used as the gate for frequency sum of S$_2$ and the LWIR pulse. In the figure, signal strength goes from red (highest) to white (lowest)}.
\label{fig:Figure 4}
\end{figure}

\begin{figure}[h!]
\centering\includegraphics[width=1.0\textwidth]{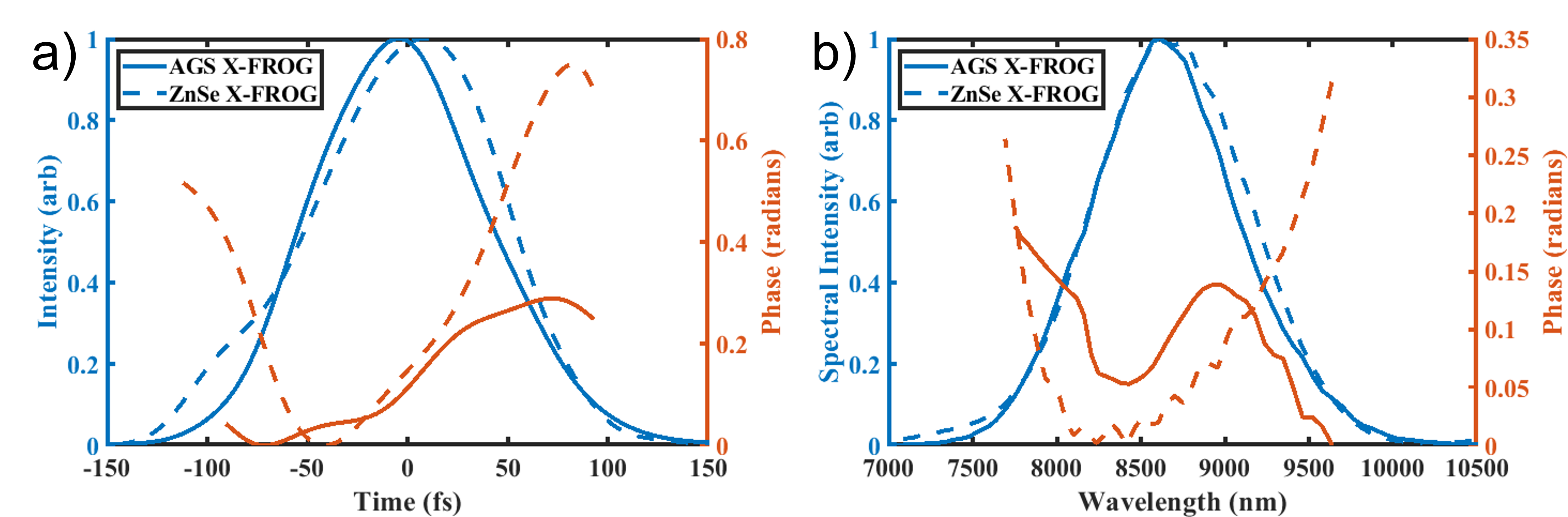}
\caption{X-FROG reconstruction of LWIR fields using both AGS and pc-ZnSe. The S$_2$ signal was used as the gate pulse and its reconstruction done with the pc-ZnSe data from above. a) reconstructed intensities and phases in the time domain , b) reconstructed intensities and phases in the spectral domain compared. In all plots blue is the normalized amplitude and orange is the phase. The rms phase errors are 0.35~rad and 0.1~rad respectively. Both X-FROG reconstructions used a grid size of 2048~x~2048. The FROG error was 0.0047 for AGS and 0.0052 for pc-ZnSe.}
\label{fig:Figure 3}
\end{figure}

Field reconstruction of the LWIR pulses using X-FROG scans were performed using S$_2$, 1.39~$\mu$m signal as the gate (Fig.\ref{fig:Figure 1}), characterized with pc-ZnSe SHG-FROG. Figure \ref{fig:Figure 4} shows raw and retrieved X-FROG spectrograms with AGS (panels a) and c)) and pcZnSe (panels b) and d)). We emphasize again that the spectrogram generated with pc-ZnSe was insensitive to the relative polarization angle between S$_2$ and LWIR whereas AGS required cross polarization, and thus produces spectrograms that are sensitive to polarization state of both fields. It should be noted that the sum-frequency intensity, while always present, did depend on the spot in pc-ZnSe the input beams were being focused on due to the different grain domains present\cite{Baudrier-Raybaut2004}. As can be seen in the raw FROG spectrograms there are some discrepancies near the temporal and spectral edges of the traces. The temporal discrepancies we believe are due to the extra accumulated phase in pc-ZnSe as compared to AGS. The spectral differences we believe are from the difference in phase matching properties between the two crystals (traditional phase matching as opposed to RQPM). In particular RQPM results in less efficient phase matching but over a broader spectral range leading to a seeming enhancement of the spectral contents near the edge. Despite these differences the retrieved spectrograms for AGS and pc-ZnSe show good agreement. A more detailed picture of is obtained by comparing full field reconstruction. Figure \ref{fig:Figure 3} shows temporal (panel a)) and spectral (panel b)) field retrievals using both AGS  and pc-ZnSe. In our opinion the match between the two methods is remarkable. Temporally we obtain a retrieved FWHM of the intensity of 103fs with  AGS and 114fs with pc-ZnSe, representing a 10\% difference, while the measured FWHM is 102.6fs with AGS and 111.1fs with pc-ZnSe. The retrieved spectral FWHM was 994nm for AGS and 1093nm for pc-ZnSe. Both crystals give the same center wavelength of 8600~nm which is fairly far from the expected DFG center wavelength between S$_1$ and S$_2$ of 9975~nm. In our experience, such large changes in center wavelength are typical in broadband DFG processes near the optical rectification regime, as the center wavelength of the generated beam is extremely sensitive to the phase matching angle of the crystal used for DFG. For the phases, the RMS deviation between the spectral and temporal phases  is 0.1~rad and 0.35~rad respectively which in our opinion also represents a very good match. At the LWIR central wavelength the group delay dispersion is approximately 100 fs$^2$ more for AGS than ZnSe and thus it could be responsible for the small deviations.

\begin{figure}[h!]
\centering\includegraphics[width=0.7\textwidth]{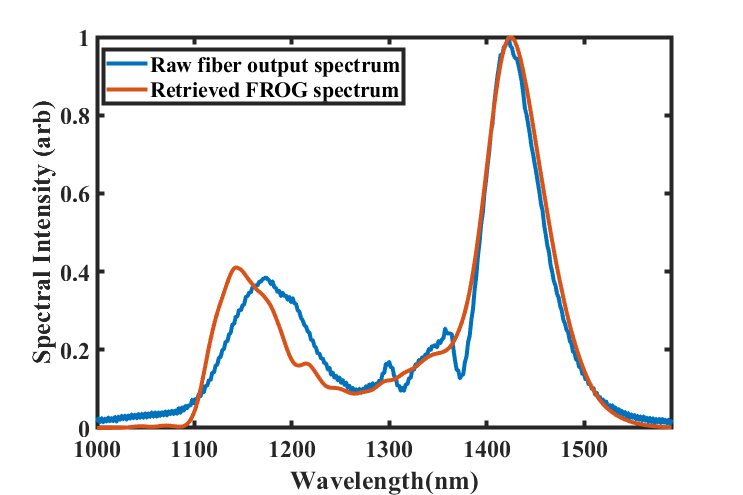}
\caption{Comparison of the raw output spectrum from the hollow-core fiber broadened Signal with the retrieved spectrum from the SHG-FROG reconstruction. As can be seen all spectra above the 7\% level is retrieved, demonstrating a large lower limit to the phase matching bandwidth of pc-ZnSe. }
\label{fig:Figure 5}
\end{figure}

The field retrievals done above yield few-cycle pulses in the LWIR with only <4 cycles under the envelope. However, the phase structure is fairly smooth since all pulses are nearly transform limited. To determine whether pc-ZnSe is capable of characterizing broadband pulses with more complicated phases, in a different experiment we broadened 800~mW of S$_2$ centered at 1.35~$\mu$m through self phase modulation (SPM). For this, we used a 1m long 300~$\mu$m diameter hollow-core fiber (Few-Cycle, Inc.) filled with argon at 30~psi. This output was collimated with a concave mirror and sent into the SHG-FROG setup with pc-ZnSe as the nonlinear medium. The beam was not compressed with chirp mirrors or any other means. As a measurement of the quality of the field reconstruction, we compared the raw measured spectrum to the retrieved one which is shown in Figure \ref{fig:Figure 5}. The broadened pulse spectrum corresponds to a Fourier limit of 10~fs or 6 cycles. As can be seen, the retrieved spectrum shows good agreement with the raw spectrum other than the pronouncement of the blue shoulder. This result demonstrates pc-ZnSe's ability to characterize broadband pulses even at a crystal thickness that would severely limit the phase matching bandwidth of most other crystals. In our opinion this result is more impressive when we consider that such an uncompressed pulse can have a complicated temporal structure due to the large amount of phase which accumulates in SPM broadened pulses.

\section{Conclusion}
In conclusion, we performed X-FROG and FROG few-cycle pulse characterization in the LWIR and NIR using one single pc-ZnSe crystal. Our experiments demonstrate pc-ZnSe's unique RQPM property to phase match second-order nonlinear processes for the purpose of SHG-FROG and X-FROG. As a result, we can characterize pulses over a decade of bandwidth independent of crystal and polarization orientation. To our knowledge no other characterization technique offers such capability. Furthermore, field retrievals show good agreement with the experimental measurement even in the case of complicated phase values as present in SPM. With its extremely large transmission window (500~nm-20~$\mu$m) pc-ZnSe is an ideal material for ultra-fast pulse characterization from the NIR all the way into the LWIR. Not only does it possess the ideal attributes of a pulse characterization crystal, a 12.5~mm diameter 1mm thick optical quality blank can be readily purchased off the shelf for under \$210, far cheaper than the assortment of crystals required for equivalent phase matching capabilities across that entire spectrum. We believe this combination of optical properties, inexpensiveness, and ease of use provides a clear path in ultra-fast pulse characterization in the MWIR to LWIR, where techniques are often more difficult and expensive. Furthermore it is interesting to note that pc-ZnSe has been shown to have a very large non-linear index of refraction \cite{Baudrier-Raybaut2004}, enabling the use of a single piece of pc-ZnSe for both second- and third-order pulse characterization over the same spectral region.

\section*{Funding}
This research was performed under the Office of Naval Research, Directed Energy Ultra-Short Pulse Laser Division grant N00014-19-1-2339. T.S. was partially funded by the US Department of Energy, Office of Science, Chemical Sciences, Geosciences, \& Biosciences Division grant DE-SC0019098.

\section*{Disclosures}
The authors declare no conflicts of interest.

\bibliographystyle{unsrt}
\bibliography{References.bib}

\end{document}